\documentclass[aps,pre,showpacs,twocolumn]{revtex4}
\usepackage{amssymb}
\usepackage{graphicx}
\usepackage{colordvi}
\usepackage{color}
\usepackage{dcolumn,amsmath,amsthm,amscd,amsfonts,amssymb,epsfig,graphics,graphicx,eucal}

\usepackage{epstopdf}

\begin{document}

\title{Bloch oscillations and resonant radiation of light propagating in arrays of nonlinear fibers with high-order dispersion}

\author{A. Yulin}
\affiliation{National Research University of Information Technologies, Mechanics and Optics (ITMO University), Saint-Petersburg 197101, Russia}

\author{R. Driben}
\affiliation{Department of Physics and CeOPP, University of Paderborn, Warburger Str. 100, D-33098 Paderborn, Germany}

\author{T. Meier}
\affiliation{Department of Physics and CeOPP, University of Paderborn, Warburger Str. 100, D-33098 Paderborn, Germany}

\date{\today}

\begin{abstract}
Bloch oscillations of spatio-temporal light wave packets in arrays of nonlinear fibers with high-order dispersion are studied.
The light wave experiences discrete spatial diffraction along the waveguide array coordinate together with continuous temporal dispersion including higher order terms.
When a gradient in the waveguide light confinement strength is considered,
the wave packet features robust long-lived Bloch oscillations with temporal and spatial spreading
in the presence of a Kerr nonlinearity.
The effect of spatial Bloch oscillations on the emission of the dispersive radiation by the solitary wave is analyzed.
It is shown that Bloch oscillations result in the generation of new frequencies. The condition of resonant emission
of dispersive waves is derived and it is demonstrated that it matches very well with the results of direct numerical simulations.
\end{abstract}

\maketitle
\input{epsf.tex} \epsfverbosetrue

\section{Introduction}

The discovery of Bloch oscillations (BOs)~\cite{Zener}
is one of the triumphs of the zone theory developed to explain the dynamics of electrons in solids in the presence of an electrical bias.
In the early 1990s, BOs were first observed experimentally in electrically-biased semiconductor superlattices
using optical interband excitation with femtosecond laser pulses~\cite{jochen}.
A few years later, BOs have been realized with atoms in optical lattices~\cite{atoms} and in optics using coupled waveguides~\cite{Peschel,Silberberg}.
It was shown in \cite{Peschel,Silberberg} that if the phase velocity of the waves changes linearly with the index $n$ of the waveguide,
the position of the light beam is an oscillating function of the propagation distance $z$ which is the optical analogue of the electronic BO dynamics.
Also the effect of a Kerr nonlinearity was considered.
However, in models with only discrete diffraction the nonlinearity has a purely destructive influence on the BO dynamics.
On the other hand, it was shown in a recent work \cite{Driben_BlochOscill} that adding another dimension with continuous diffraction may result
in a constructive effect of the nonlinearity, i.e., a localization of the wave packet in space
and the creation of a quasi-solitonic regime of propagation.

Anomalous dispersion in time can act together with the discrete diffraction and self-focusing nonlinearity towards the creation of robust spatio-temporal localized wave-packets
experiencing BOs.
Moreover, considering high-order temporal dispersion suggests the investigation of radiative properties of the robust nonlinear wave packets oscillating with a period defined by BOs.
This is the core of the results reported in this paper.
It is natural to compare the dynamics of the pulse in the presence of BOs with the propagation of the pulse in the perfectly periodic discrete-continuous system without an
refractive index gradient which is provided in Appendix~\ref{appb}.

The problem of resonant radiation of dispersive waves by optical solitons has been studied for long times.
This problem is of significant practical importance for example for optical supercontinuum generation \cite{Skryabin_RMPhys} and, at the same time,
is an interesting fundamental problem for nonlinear waves dynamics.
It has been shown that in the presence of high-order dispersion (HOD) the solitons start emitting radiation because of Cherenkov synchronism of the solitons with the dispersive waves \cite{Menyuk,Akhmediev}.
In periodic systems this phenomenon can be interpreted as transitional radiation or Cherenkov radiation of Bloch waves \cite{Yulin_Transitional}.
The recoil from the resonant radiation modifies the properties of the solitons \cite{Skryabin_Science,Bianca_recoil} and affects the frequency of the soliton and the resonant radiation.
A phenomenon closely related to Cherenkov radiation and very relevant for our present studies is the radiation of the dispersive waves by oscillating solitary structures such as high-order solitons or solitons in fibers with dispersion management that were investigated both theoretically \cite{Theory} and experimentally \cite{Experiment}.
Also spectral lines with similar characteristics were observed in the propagation of spatio-temporal
oscillating nonlinear waves in multimode fibers \cite{multimode}.
In the present paper we consider the dynamics of light in the framework of a conservative scalar model which is applicable when modes of different polarizations are detuned significantly
and therefore do not interact strongly.
However, we would like to mention that the discussed phenomena can possibly also be observed in vector and active (lasing) systems where complex solitary structures
may form, see for example \cite{VecSolLas,DomWallLas}.

To describe the evolution of the field we adopt the model used in \cite{Aceves}.
The system is continuous in time $t$, discrete along the coordinate $n$,
and the evolution coordinate is the propagation distance $z$.
The propagation constant measured at the pump frequency depends linearly on the index of the waveguide $n$,
whereas the group velocity and the dispersion are the same for all the waveguides.
The discrete-continuous model is very popular due to the richness of nonlinear phenomena, see, e.g., \cite{Overview} and references within.
Mathematically, such a system is described by the equation
\begin{eqnarray}
-i\partial_z \psi_n = \hat D \psi_n + \sigma( \psi_{n+1}+\psi_{n-1}-2\psi_{n})+ \nonumber \\ +\alpha |\psi_{n}|^2\psi_{n} +\gamma n \psi_{n}  , \label{m_eq1}
\end{eqnarray}
where $\hat D$ is an operator describing the dispersion of the pulse.
Here, we follow the practice adopted in the optics community and work in the relative reference frame and
use the retarded time $t$ as an evolution coordinate and provide in Appendix~\ref{appa}
the transformation from the usual description of light wave packet dynamics to Eq.~(\ref{m_eq1}).

We consider here the simplest case of a quadratic dispersion with $\hat D=\frac{1}{2}\partial_t^2$
and the case of relatively strong third order dispersion $\hat D=\frac{1}{2}\partial_t^2+i\frac{1}{3}\beta_3 \partial_t^3$.
It should be mentioned that the results can easily be generalized to the case of more complicated HOD.
In this work, we restrict the analysis to the case of relatively long pulses for which the Raman effect can be neglected.
All simulations presented below have been performed using the following normalized parameters of the fiber
$\sigma=10$, $\gamma=5$, and $\alpha=0.2$.

In the numerical simulations the field distribution of the initial pulse is taken as
\begin{eqnarray}
\psi_n(t, z=0)=a_0 e^{-(t-t_0)^2/w_t^2-n^2/w_n^2} e^{i k_0 n} , \label{ini_cond}
\end{eqnarray}
where $a_0$ is the amplitude, $w_t$ and $w_n$ are the widths of the pulse along $t$ and $n$ coordinates correspondingly, and $t_0$ is the shift of the pulse along $t$.
The shift $t_0$ is introduced to provide better accommodation of the field in the simulation window.
The parameters of the initial pulse are chosen as $a_0=1$, $w_t=5$, $w_n=40$, and $k_0=0$.
We would like to point out that the qualitative behavior of the system does not depend on the precise values of the fiber and initial pulse parameters
provided that they allow the formation of a reasonably frequently oscillating wavepacket.

\section{Bloch oscillations of nonlinear wave packets and radiation emission without high-order dispersion}

\begin{figure}
  \includegraphics[width=0.5\textwidth]{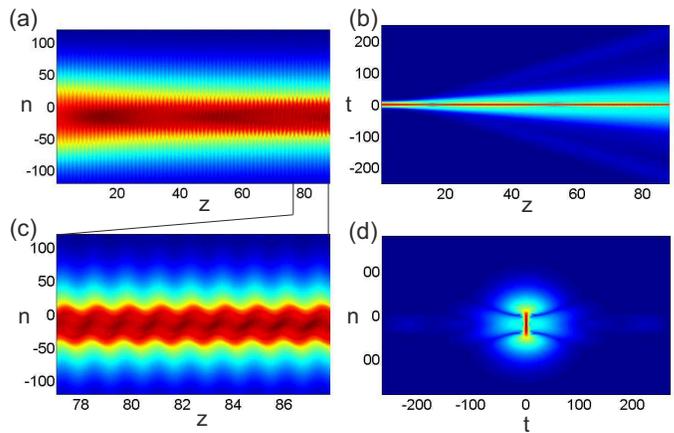}
  \caption{(Color online) The dynamics of the field for quadratic dispersion ($\beta_3=0$). The initial time shift of the pulse is $t_0=0$. The evolutions of $|\psi_n(t=0)|$ and $|\psi_{n=0}(t)|$ as function of the propagation distance $z$ are shown in panels (a) and (b) correspondingly. Zoomed quasi-stationary oscillations of $|\psi_n(t=0)|$ are illustrated in panel (c). Panel (d) shows the field distribution $|\psi_n(t)|$ at $z=88$.}\label{Fig1}
\end{figure}

We start with the simple case of pure quadratic dispersion.
Results of numerical simulations illustrating the typical dynamics of the system are presented in Fig.~\ref{Fig1}.
Quasi-periodic BOs along the discrete coordinate-$n$ are clearly seen for long propagation distances in panel (a),
with radiation shedding as well seen in the temporal domain in panel (b).
The field distribution after relatively long propagation distance starting from $z=77$ is shown in panel (c),
while the field profile in the $n$-$t$ plane is displayed in panel (d).
The oscillations in the nonlinear case are noticeably asymmetric with respect to $n$, best seen in panel (c),
however, the field still remains well localized \cite{Driben_BlochOscill}.
At the same time it is seen in panel (b) that the pulse stays relatively well localized along the temporal coordinate $t$.
However, weak radiation is shedding symmetrically away from the main part of the pulse along the $t$ direction.
The shape of the area filled with the radiation supports the conclusion that the field stays mostly localized
but there is weak leakage along $t$ coordinate.

\begin{figure}
  \includegraphics[width=0.35\textwidth]{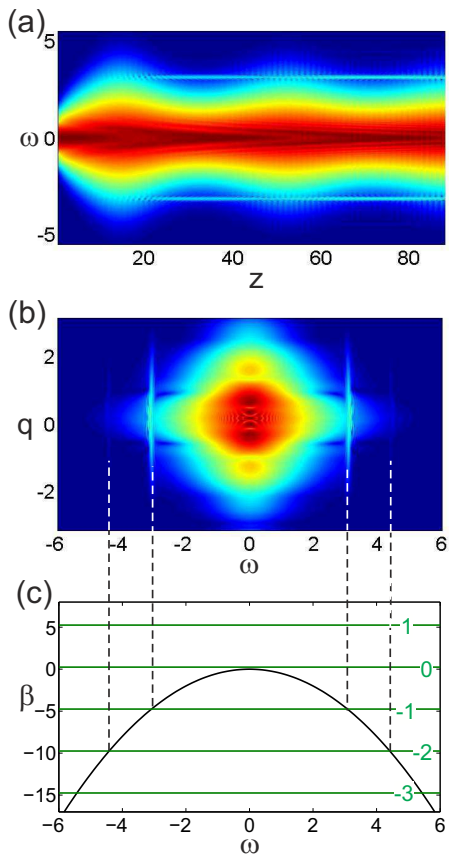}
  \caption{(Color online) The dynamics of the spectrum of $\psi_{n_c}(t)$ as function of the propagation distance $z$ is shown in panel (a).
At each propagation distance, $n_c(z)$ corresponds to the central point of the pulse and is defined as $n_c(z)=  \left \lceil{  \int \sum_n( |\psi_n(t)|^2 n) dt/\int \sum_n |\psi_n(t)|^2  dt) -\frac{1}{2}}\right \rceil$.
The spatial-temporal spectrum of the field $\psi_n(t)$ at $z=88$ is shown in panel (b).
$q$ is spatial wave vector corresponding to the $n$-axis.
A graphical solution of the resonance condition, Eq.~(\ref{res_cond_prlm}), with $K_s=0.25$ is given in panel (c) where the value of $K_s$ is an estimate based on the numerical simulations.
The horizontal green lines correspond to the resonances between different Bloch harmonics of the solitary pulse and the radiated wave.}
The spectra are calculated for the fields illustrated in Fig.~\ref{Fig1}.
\label{Fig2}
\end{figure}

It is instructive to examine the dynamics of the field in the spectral representation.
The evolution of the spectrum of the field section is shown in panel (a) of Fig.~\ref{Fig2} where it is seen that during the propagation
the emission of dispersive waves symmetric spectral lines forms.
These spectral lines are better seen in the spatial-temporal spectrum of the field at the propagation distance $z=88$, see panel (b).
The first two lines are quite pronounced and the second pair of the lines are of much lower intensity but still visible.
It should be noted here that for the considered case the width of the first and the second radiation lines along $q$ are approximately the same.
The reason why the second spectral line may look narrower than the first one is that the low intensity of the second line results
in poor resolution of the wings of the line.

In the next section we will demonstrate that these lines are associated with the resonant radiation emitted by Bloch oscillating wave packets.

\section{Resonance condition}

To understand the formation of the radiation, we use a perturbative approach to derive the condition for the resonant emission of dispersive waves by the solitary structure.
The derivation presented below is very similar to previous studies on the resonant emission of waves by oscillating solitons \cite{Theory}.
We assume that a solution has the form of a localized oscillating structure $\tilde \psi$
and a small correction accounting for the radiation field $\varphi$.
The equation for $\varphi$ takes the form
\begin{eqnarray}
i\partial_z \varphi_n +\hat D \varphi_n + \sigma( \varphi_{n+1}+\varphi_{n-1}-2\varphi_{n})+ \nonumber \\ +2 \alpha |\tilde \psi_{n}|^2 \varphi_{n}+\alpha \tilde \psi_{n}^2 \varphi_{n}^{*} +\gamma n \varphi_{n} =f   ,
\label{m_eq2}
\end{eqnarray}
where
\begin{eqnarray}
 f=-i\partial_z \tilde\psi_n -\hat D \tilde \psi_n - \sigma( \tilde \psi_{n+1}+\tilde \psi_{n-1}-2\tilde \psi_{n})-
\nonumber \\ -\alpha |\tilde\psi_{n}|^2\tilde \psi_{n} -\gamma n \tilde \psi_{n} \label{m_eq2_2}
\end{eqnarray}
is the source.

The localized component of the field $\psi$ experiencing BOs  can be represented as
$$\tilde \psi_n= \sum_{\tilde m} g_{\tilde m} e^{i \tilde m K_b z} e^{i \tilde K_s z}$$
where $g_{\tilde m}$ are localized functions of $n$ and $t$ describing the field distributions in the harmonics of BOs,
$\tilde m$ numerates the harmonics, $K_b$ is the Bloch frequency, and $\tilde K_s$ is the soliton propagation constant.
The spatial frequency of BOs can be controlled by the parameter $\gamma$ and for our model the relation $K_b=\gamma$ holds, see \cite{Driben_BlochOscill}.

It is important to note that in the oscillating solitary waves all harmonics move as a whole with the same velocity.
Therefore, the functions $g_{\tilde m}$ are actually functions of only two arguments $n$ and $t-\mu_s z$, so we have $g_{\tilde m}=g_{\tilde m}(n, t-\mu_s z)$.
Without loss of generality we can consider the soliton resting in the chosen reference frame and set $\mu_s=0$.
Then expanding each function $g_{\tilde m}$ in Fourier series over the second argument we arrive at the following representation of $\tilde \psi$
\begin{eqnarray}
\tilde \psi_n= \sum_{\tilde m} \int_\omega h_{\tilde m}(n, \omega) e^{-i\omega t} e^{i \tilde m K_b z} e^{i \tilde K_s z}.
\label{tilde_psi_Four}
\end{eqnarray}
Substituting Eq.~(\ref{tilde_psi_Four}) into Eq.~(\ref{m_eq2_2}) we see that  the source term can be written as
\begin{eqnarray}
f= \sum_{\tilde m} \int_{\omega} f_{\tilde m}(n, \omega) e^{-i\omega t} e^{i \tilde m K_b z} e^{i \tilde K_s z}.
\label{source}
\end{eqnarray}

Now let us consider the eigenmodes of the medium. The equation for free propagating linear waves is
$$i\partial_z \varphi_n +\hat D \varphi_n + \sigma( \varphi_{n+1}+\varphi_{n-1}-2\varphi_{n}) +\gamma n \varphi_{n}=0$$
and from here it is seen that the eigenmodes can be sought in the form
$\varphi=\phi(n, z)e^{i\beta z-i\omega t}$,  where $\beta(\omega)$ is the Fourier representation of the operator $\hat D$.
The physical meaning of $\beta$ is the dispersion of linear waves in the central waveguide with $n=0$.
The equation for $\phi$ reads
\begin{eqnarray}
i\partial_z \phi_n + \sigma( \phi_{n+1}+\phi_{n-1}-2\phi_{n}) +\gamma n \phi_{n} =0.
\label{eif}
\end{eqnarray}

The eigenmodes of the latter equation are localized in $n$ and oscillate with the Bloch frequency. The eigenfunction with the propagation constant $K_w$ has the form $h=\sum_m c_{m} h_m(n)e^{i (m K_b+K_w) z}$,
where $c_{m}$ are the coefficients of the expansion and $h_m$ are localized functions describing the transverse structure of different harmonics of the mode.
The symmetry of Eq.~(\ref{eif}) provides that if $h(n)$ is a solution then $h(n+n_0)e^{i \gamma n_0 z}$ is a solution too ($n_0$ is an integer).
Therefore, the eigenmodes can be parametrized by their propagation constant and by the position of their center $n_0$.

Finally, we come to the conclusion that the eigenmodes of the systems can be written as
\begin{eqnarray}
\varphi=e^{i(K_w + \gamma n_0 +\beta) z-i \omega t}\sum_m c_{m} h_m(n-n_0)e^{i m K_b z}.
\label{lin_fld_exp}
\end{eqnarray}
The resonant emission occurs if one of the harmonics of the eigenmode, Eq.~(\ref{lin_fld_exp}),
is phase matched with a harmonic of the source, Eq.~(\ref{source}). This results in the resonance condition
$$ K_w + \gamma n_0 +\beta+ m K_b=lK_b+\tilde K_s. $$
Taking into account that $K_b=\gamma$ and introducing the detuning of the soliton propagation constant from the propagation constant of the linear mode $K_s=\tilde K_s-K_w$,
we arrive at the resonance condition
\begin{eqnarray}
\beta(\omega)=lK_b+K_s,
\label{res_cond_prlm}
\end{eqnarray}
where $l$ is an integer.
From the resonance condition it is clear that the generated frequencies can be controlled by the dispersion properties of the system and by the period of BOs.
Also the intensity of the initial pulse affects the resonance frequencies but typically this correction is negligible.

Let us compare how well the resonance condition, Eq.~(\ref{res_cond_prlm}),
predicts the positions of the spectral lines generated by the solitary structures experiencing BOs.
A graphical solution of the resonance condition is given in panel (c) of Fig.~\ref{Fig2}, the parameter $K_s$ was extracted from the numerical simulations.
It is seen that the positions of the spectral lines observed in numerical simulations match very well the resonant frequencies given by
Eq.~(\ref{res_cond_prlm}). The symmetry of the system ensures that the radiation is emitted symmetrically in both directions.
This is why in this case the radiation cannot change the velocity of the solitary wave or break its symmetry.

\section{Resonant radiation emission in the presence of high-order dispersion}

\begin{figure}
  \includegraphics[width=0.5\textwidth]{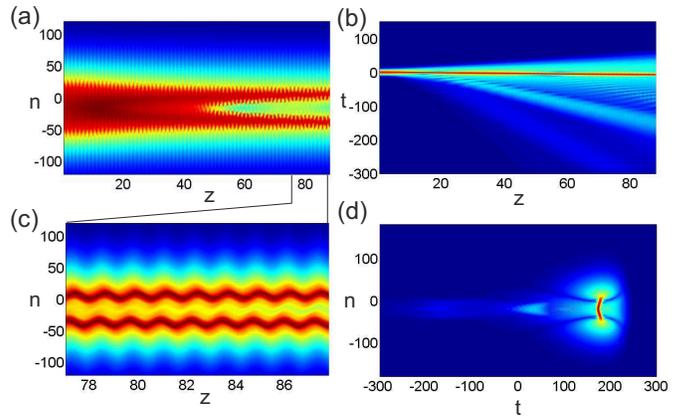}
  \caption{(Color online) The same as Fig.~\ref{Fig1} but for $\beta_3=0.45$ and an initial time shift of $t_0=180$.}\label{Fig3}
\end{figure}

Now we study how HOD affects the resonant radiation. Relatively small HOD does not destroy the solitary pulse, see panels (a), (b), and (c) of Fig.~\ref{Fig3}
which illustrates the evolution of the pulse in the presence of HOD.
The localization and BOs are well preserved in the presence of HOD but it is seen that the symmetry is now broken.
We would like to point out that in the case of third order dispersion the sign of $\beta_3$ defines only whether the frequency of the Cherenkov synchronism
is higher or lower than the frequency of the solitary pulse. That is why we discuss in this paper only the case with positive $\beta_3$.

Panel (d) of Fig.~\ref{Fig3} shows the distribution of the field at a long propagation distance $z=88$. Firstly, it is seen that the radiation tail is emitted predominantly in one direction.
It is worth mentioning that the radiation tail remains localized along the discrete coordinate $n$ and experiences BOs.
Secondly, the shape of the localized part of the pulse is deformed. As seen in panel (d) of Fig.~\ref{Fig3}, the center of the solitary wave is delayed with respect to its peripheral area.
This deformation explains why in the propagation of the section of the field at $t=0$ two maxima develop, see panels (a) and (b) of Fig.~\ref{Fig3}.
This behavior with HOD differs from the case of a pure quadratic dispersion where the field stays symmetric during propagation, see Fig.~\ref{Fig1}.
Let us note that the recoil from the radiation can be one of the reasons which causes the deformation of the localized part of the field.

\begin{figure}
  \includegraphics[width=0.35\textwidth]{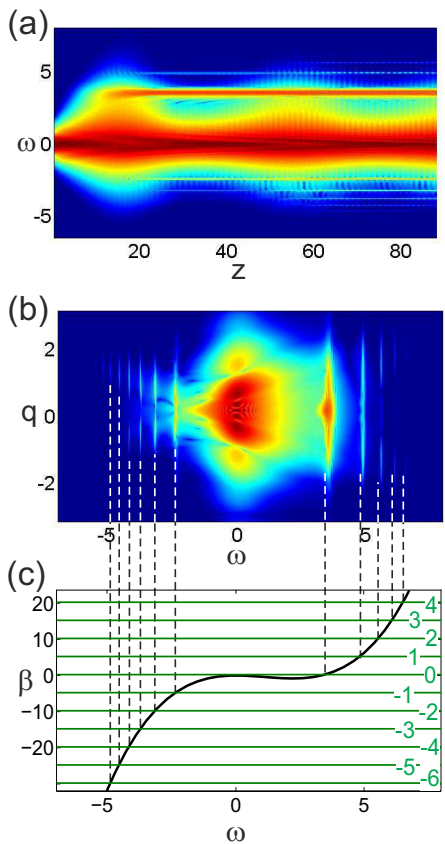}
  \caption{(Color online) The same as Fig.~\ref{Fig2} but for $\beta_3=0.45$ and an initial time shift of $t_0=180$.}\label{Fig4}
\end{figure}

Now let us turn to the spectrum of the field propagating in the system with HOD. The evolution of the spectrum is show in panel (a) of Fig.~\ref{Fig4}.
It is seen that an asymmetric set of spectral lines grows during propagation.
The spatial-temporal spectrum of the field at a long propagation distance is shown in panel (b) and it is also asymmetric.
The graphical solution of the resonance condition, Eq.~(\ref{res_cond_prlm}), is presented in panel (c) for the parameter $K_s$ extracted from the direct numerical simulations.
The comparison of the positions of the spectral lines obtained from the numerical simulations with the prediction of the resonance condition shows that the observed spectral lines can indeed
be attributed to the resonant emission of dispersive waves by a main part of the wave experiencing BOs.
The resonance with $l=0$ corresponds to Cherenkov radiation and this radiation channel is not much affected by the BOs.
The other spectral lines cannot occur without BOs. This is why BOs of the quasi-soliton significantly enrich the spectrum of the radiated dispersive waves.
It is worth to notice that such a frequency peak train radiation is typical for periodically-oscillating nonlinear objects emitting resonant radiation in the course of propagation.
Examples for other types of periodically-oscillating robust nonlinear waves have been presented in \cite{Theory,Experiment}.

In the absence of HOD, only resonances corresponding to negative $l$ in Eq.~(\ref{res_cond_prlm}) exist.
With HOD the resonances with zero and positive $l$ become possible and which is an important difference between the systems with and without HOD.
Let us consider the specific example illustrated by Fig.~\ref{Fig4}.
The radiation with frequencies lower than the frequency of the quasi-solitary wave appears due to the resonances with negative $l$ and
thus this radiation is generated by the same resonances as the resonant radiation in the system without HOD.
However, the radiation with frequencies higher than the soliton frequency is very different.
It is generated by the resonance with positive $l$ meaning that this radiation is due to different spectral components of the solitary wave and thus has a different physical origin.
It is therefore possible to state that HOD results in the Cherenkov synchronism and BOs produce additional Cherenkov-like lines.

\begin{figure}
  \includegraphics[width=0.35\textwidth]{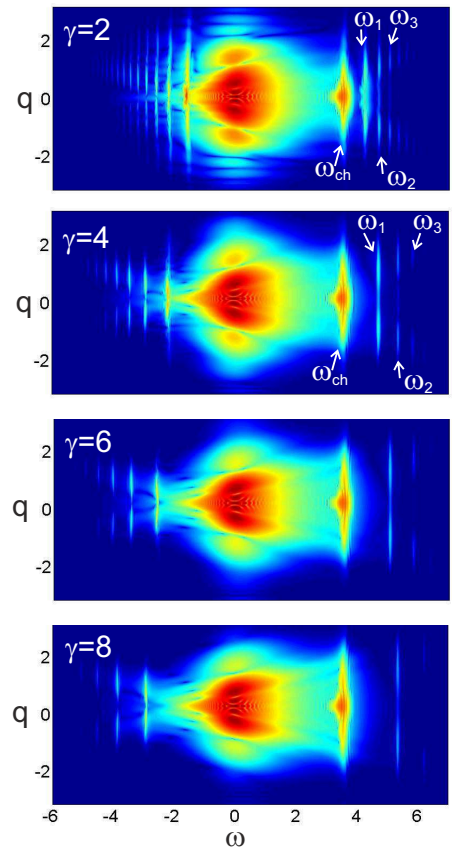}
  \caption{(Color online) The spatial-temporal spectra of the field at the propagation distance $z=88$  for different gradient strengths $\gamma$.}\label{Fig4_1}
\end{figure}

To provide additional evidence that the resonant lines appear because of BOs we study how the spectrum of the field at long propagation distance depends of the period of BOs.
The spectra of the field at $z=88$  are shown in Fig.~\ref{Fig4_1}.
It is clearly seen that the positions of the resonance lines depend on the Bloch frequency.

\begin{figure}
  \includegraphics[width=0.45\textwidth]{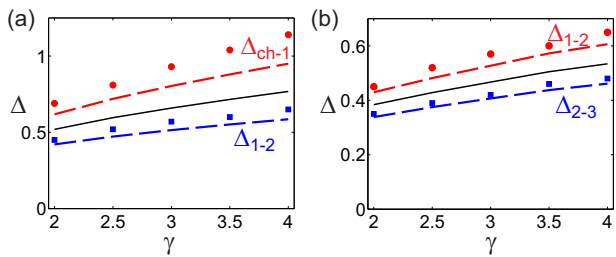}
  \caption{(Color online) The dependencies of the spectral distances between the neighboring lines as functions of the parameter $\gamma$ defining the spatial frequency of the BOs.
  Panel (a) corresponds to the spectral line marked as $\omega_1$ in Fig.~\ref{Fig4_1}.
  The dark solid line corresponds to the first-order approximation for the distance between neighboring lines. The red dashed line shows the second-order approximation for the distance to the neighboring line having lower frequency (the line is the Cherenkov radiation line and marked as $\omega_{ch-1}$ in Fig.~\ref{Fig4_1}). The spectral distances extracted from direct numerical simulations are shown by red circles. The blue dashed line marked as $\Delta_{1-2}$ shows the second-order approximation for the distance to the neighboring spectral line having higher frequency (the line is marked as $\omega_2$ in in Fig.~\ref{Fig4_1}). Panel (b) shows the same but for the line marked as $\omega_2$ in Fig.~\ref{Fig4_1}. The curve $\Delta_{1-2}$ and the circles are for the spectral distance to the neighboring line having lower frequency and the blue curve $\Delta_{2-3}$ and the blue circles  are for the neighboring spectral line having higher frequency (the line is marked as $\omega_3$ in Fig.~\ref{Fig4_1}).}\label{Fig4_2}
\end{figure}

The resonance condition Eq.~(\ref{res_cond_prlm}) allows to obtain a simple estimate for the difference between neighboring spectral lines.
The first-order approximation for the spectral distance to the neighboring line is
\begin{eqnarray}
\Delta^{(1)}=\left|  \frac{K_b}{\mu_1} \right|,
\label{sp_dist_1}
\end{eqnarray}
where $\mu_1=\partial_{\omega} \beta$ is calculated at the frequency of the line.
The second-order approximation gives
\begin{eqnarray}
\Delta^{(2)}=\left|  \frac{K_b}{\mu_1} \right| (1 \mp \frac{\mu_2 K_b}{\mu_1^2} ),
\label{sp_dist_2}
\end{eqnarray}
where $\mu_2=\partial_{\omega}^2 \beta$.
The sign $+$ should be taken for the distance to the line with higher frequency and the sign $-$
has to be taken for the distance to next lower frequency line.
The comparison between the numerical data and the analytical results, Eq.~(\ref{sp_dist_1}) and Eq.~(\ref{sp_dist_2}), are shown in Fig.~\ref{Fig4_2}.
The agreement is very good and clearly better for smaller values of $K_b$ when the distances between the lines are smaller.
This result is to be expected because the expressions of Eq.~(\ref{sp_dist_1}) and Eq.~(\ref{sp_dist_2}) are obtained by an expansion of the resonance condition
into a Taylor series assuming that $\Delta$ is small.

Drawing an analogy with the radiation emitted by moving charges, the emission of waves by BOs of solitary waves corresponds to the emission of the electromagnetic waves by oscillating dipoles moving at a subluminal
velocity. The case with HOD is analogous to the radiation of oscillating dipoles moving at a superluminal velocity and thus in this case the resonant radiation contains the Cherenkov component.
Since the resonant radiation introduces effective losses for the solitary wave and the soliton energy decreases during the propagation.
However, for the chosen parameters the losses do not strongly affect the dynamics of the solitary waves.
The case of stronger HOD is important but beyond the scope of the present paper.

\section{Conclusion}

In this paper, we demonstrate that Bloch oscillations of solitary waves propagating in nonlinear optical fiber arrays cause multi-modal resonant radiation which is somewhat analogous to the radiation of oscillating dipoles.
The resonance condition is derived and it is shown that the predicted resonances are in excellent agreement with the positions of the spectral lines observed in numerical simulations.
In the presence of high-order dispersion one of the spectral lines corresponds to Cherenkov radiation and this can be interpreted as the radiation generated by the zeroth harmonic of the Bloch oscillations.

The numerical simulations show that the propagation of optical solitons in the regime of Bloch oscillations can be robust and that the generated multi-mode radiation of the dispersive waves should be observable experimentally.
The understanding of the phenomenon of resonant radiation in the presence of Bloch oscillations can be useful for the optical supercontinuum generation in the arrays of optical fibers.
The role of effects like Raman and self-steepening could be important for the propagation of ultra-short pulses, however, this is beyond the scope of the present paper where we consider long pulses
and will be considered elsewhere.

We believe that the discussed phenomenon of resonant emission of frequency combs by a solitary wave experiencing Bloch oscillations can be observed in systems similar to the
waveguide arrays studied in \cite{Silberberg}. To facilitate the observation, the frequency of the initial optical pulse should be sufficiently close to the zero dispersion point which can be obtained
by adjusting the parameters of the waveguides.
Another system for the possible realization of resonant emission of dispersive waves by oscillation spatial-temporal solitary wave
are multicore silica fibers.
We believe that potentially these systems can be used for example for supercontinuum generation.
It is worth mentioning that in the discussed systems the parameters of the supercontinuum should depend on the position
where the initial pulse enters the waveguide array.

\section*{Acknowledgements}
The authors acknowledge fruitful discussions with Vladimir Konotop.
The work of AY was supported by the Government of the Russian Federation (Grant 074-U01) through the ITMO University early career fellowship.
T.M. and R.D. acknowledge support of the DFG (Deutsche Forschungsgemeinschaft) through the TRR 142 (project C02) and thank the PC$^2$ (Paderborn Center
for Parallel Computing) for providing computing time.

\appendix
\section{Transformation from the light wave packet dynamics description in terms of physical units to Eq.~(1)}
\label{appa}

The propagation of laser light pulse in media with a Kerr-type nonlinearity and a refractive index variation can be described by the following equation:
\begin{eqnarray}
2ik_{0}A_{z} +A_{zz} +\sum _{m =1}^{\infty }\frac{1}{m !}\frac{ \partial ^{m}}{ \partial \omega ^{m}}(\imath \partial _{t})^{m}A +\nonumber \\ +\frac{ \partial ^{2}}{ \partial x^{2}}A +\frac{4n_{0}\omega _{0}^{2}}{c^{2}}(n_{2}\vert A\vert ^{2} +\Delta n)A =0 ,
\end{eqnarray}
where $A$ is the slowly varying amplitude and $k^{2}(\omega) =\omega^{2}n_{0}^{2}/c^{2}$, with $k_{0}=k(\omega _{0})$.
Following the analysis described for example in \cite{Fibich} we change to the moving-frame coordinate system $\tilde{x}=x/r_{0}, \tilde{z}=z/2L_{diff}, \tilde{t}=(t -z/c_{g})/T_{0}$ ,
and $\psi=r_{0}k_{0}\sqrt{4n_{2}/n_{0}}A$. Here, $r_{0}$ is the spatial wave packet width, $L_{diff}=r_{0}^{2}k_{0}$ is the diffraction length, $T$ is the pulse duration, and $c_{g}= 1/k'(\omega _{0})$ is the group velocity. When the pulse duration is sufficiently short the following transformation can be made with dropping the tildes:
\begin{eqnarray}
-i\partial_z \psi = \hat D \psi +\psi_{xx} + |\psi|^2\psi +\gamma x \psi  . \label{m_eq01}
\end{eqnarray}

The parameter $\gamma$$x$ is proportional to $\Delta n/n_2$ and describes the particular linear gradient refractive index variation across the dimension $x$. Considering the discrete nature of the spatial dimension of our model the diffraction term with its second derivative can be replaced by its discrete counterpart with $\sigma$ representing the strength of coupling between the neighbor waveguides. Thus we replace the continuous coordinate $x$ by the discrete coordinate $n$ describing the order of waveguides in a array. We can also control the strength of nonlinear term by an additional coefficient $\alpha$ that is proportional to the intensity of the input light.
Hereafter we add this coefficient. Following the outlined procedure, the above given Eq. (\ref{m_eq01}) is replaced by its discrete-continuous counterpart:
\begin{eqnarray}
-i\partial_z \psi_n = \hat D \psi_n + \sigma( \psi_{n+1}+\psi_{n-1}-2\psi_{n})+ \nonumber \\ +\alpha |\psi_{n}|^2\psi_{n} +\gamma n \psi_{n}  .
\end{eqnarray}

Let us take a closer look at the operator describing the temporal dispersion of the pulse, i.e., $\hat D$.
After the vanishing of the first order derivative due to changing the reference frame we further take into account the two leading terms namely the quadratic and the cubic terms.
Thus $\hat D=-d_{2}\psi_{tt}-id_{3}\psi_{ttt}$, where $d_{2}=L_{diff}/L_{disp}$, with $L_{disp}=T^2/k''(\omega _{0})$ and $d_{3}=r_{0}^{2}k(\omega _{0})k'''(\omega _{0})/(3T^3)$.
In the anomalous dispersion regime $d_{2}$ is negative and for simplicity we continue our discussion by setting the quadratic coefficient to $d_{2}=-0.5$ and introducing $\beta_{3}=-d_{3}/3$.
Thus our dispersion operator is  slightly simplified into $\hat D=\frac{1}{2}\partial_t^2+i\frac{1}{3}\beta_3 \partial_t^3$.

\section{Resonant radiation emission without gradient in the presence of high-order dispersion}
\label{appb}

\begin{figure}
  \includegraphics[width=0.45\textwidth]{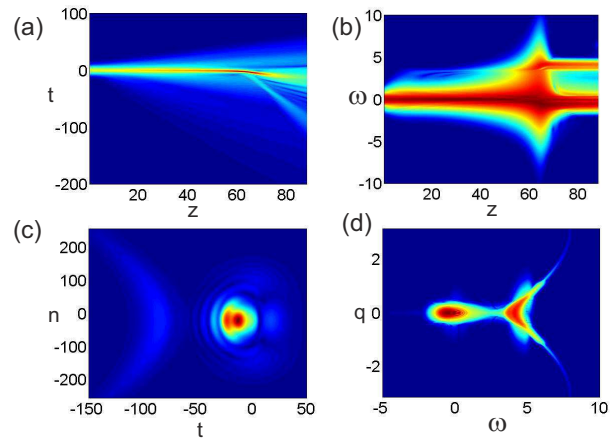}
  \caption{(Color online) The evolution of $\psi_{n=n_c}(t)$ along time is shown in panel (a). Panel (b) shows the evolution of the temporal spectrum of $\psi_{n=n_c}(t)$.
  The field intensity and the spatial-temporal spectrum are shown in panels (c) and (d) for a propagation distance of $z=88$.
  The third-order dispersion is $\beta_3=0.45$, the wave vector of the initial envelope is $k_0=0$, and the amplitude of the initial pulse is $a_0=0.7$.}\label{Fig4_3}
\end{figure}

For sake of completeness and for comparison we show here how the radiation process takes place in the presence of HOD but without Bloch oscillations, i.e., for $\gamma=0$.
The propagation of the envelope with central wavevector $k_0=0$ is shown in Fig.~\ref{Fig4_3}. As discussed above, without HOD, i.e., $\beta_3=0$, the pulse is getting narrow and the collapse starts to develop. Panel (a) of Fig.~\ref{Fig4_3} shows that at a short propagation distance the initial pulse gets compressed in the presence of HOD too.
However, in this case the compression of the pulse is accompanied by the emission of dispersive waves. The emission rate grows very rapidly when the width of the pulse decreases and, consequently, the pulse becomes spectrally wide. The broadening of the spectrum is seen in panel (b) as well. At a certain propagation distance the radiation becomes very strong and Cherenkov lines form in the spectrum.
Due to this strong radiation the pulse is loosing energy and thus this process is an alternative mechanism of the arrest of the collapse in two-dimensional optical waveguides with a Kerr nonlinearity \cite{Yulin2D}.

The distribution of the field and the spectrum of the field at long propagation distance are shown in panels (c) and (d) of Fig.~\ref{Fig4_3}. It is seen that a radiation cone forms so that different frequencies propagate at different angles with respect to the propagation direction of the emitting pulse. This is very similar to what happens in spatially uniform systems \cite{Yulin2D}, however, the discreteness of one of the coordinates affects the dispersion characteristics and, consequently, the shape of the radiation cone. A detailed study of this phenomenon is beyond the scope of the present paper.
What is important here is that in the absence of BOs only the Cherenkov radiation can be observed.
Another important aspect is that in the presence of BOs the radiation does not form a cone but propagates along the fibers.

\begin{figure}
  \includegraphics[width=0.45\textwidth]{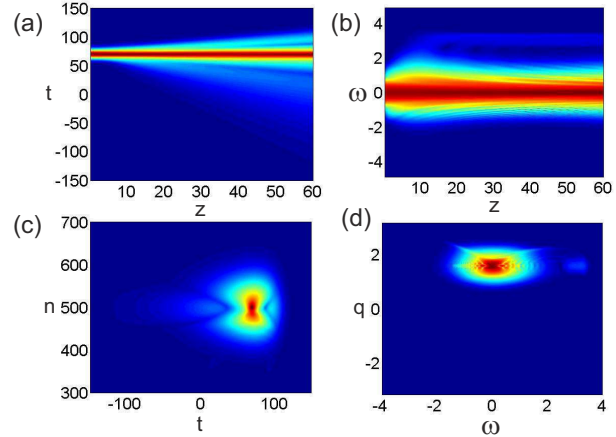}
  \caption{(Color online) The same as Fig.~\ref{Fig4_3} but for an initial field distribution with $k_0=\pi/2$.}\label{Fig4_4}
\end{figure}

A last remark that needs to be made is that the tendency towards collapse is not a necessary condition for the Cherenkov radiation.
In the presence of HOD, Cherenkov radiation can be observed for example in the course of propagation of a pulse forming from the initial condition with $k_0=\pi/2$.
This is illustrated in Fig.~\ref{Fig4_4} which shows the evolution of the pulse and the distributions of the field and the spectral intensities at relatively long propagation distances.
It should be noted that there is a frequency emitted with the highest efficiency but the entire range of frequencies is radiated.
For the purpose of this paper it is important that only Cherenkov radiation takes place.

\end{document}